\documentclass[letterpaper]{article}
\usepackage{aaai}
\usepackage{times}
\usepackage{helvet}
\usepackage{courier}
\usepackage{graphicx}
\usepackage{subfigure}
\usepackage{epsfig}
\usepackage{amssymb} \usepackage[latin1]{inputenc}
\usepackage[usenames]{color}
\usepackage{paralist}
\usepackage{amsmath}
\usepackage{balance}

\begin{document}
%
\title{What is Tumblr: A Statistical Overview and Comparison} 
\author{
Yi Chang\textsuperscript{\ddag}, Lei Tang\textsuperscript{\S}, Yoshiyuki Inagaki\textsuperscript{\dag} \and Yan Liu\textsuperscript{\ddag}\\
\textsuperscript{\dag} Yahoo Labs, Sunnyvale, CA 94089, USA\\ 
\textsuperscript{\S} @WalmartLabs, San Bruno, CA 94066, USA\\ 
\textsuperscript{\ddag} University of Southern California, Los Angeles, CA 90089\\ %
yichang@yahoo-inc.com,leitang@acm.org, \\inagakiy@yahoo-inc.com,yanliu.cs@usc.edu
}
\maketitle

\begin{abstract}
  Tumblr, as one of the most popular microblogging platforms, has
  gained momentum recently.  It is reported to have 166.4
  millions of users and 73.4 billions of posts by January 2014. While many articles
  about Tumblr have been published in major press, there is not much
  scholar work so far. In this paper, we provide some pioneer
  analysis on Tumblr from a variety of aspects.  We study the social network structure among
  Tumblr users, analyze its user generated content,
  and describe reblogging patterns to analyze its user behavior.   We aim
  to provide a comprehensive statistical overview of
  Tumblr and compare it with other popular social services, including
  blogosphere, Twitter and Facebook, in answering a couple of key
  questions: \emph{What is Tumblr?  How is Tumblr different from other
    social media networks?}  In short, we find
  Tumblr has more rich content than other microblogging platforms, and it
  contains hybrid characteristics of social networking, traditional
  blogosphere,  and social media.
This work serves as an early snapshot of Tumblr that later work can leverage.\footnote{This paper will be officially published on SIGKDD Explorations 2014, please cite official version.}\footnote{This work is also reported by MIT Technology Review at http://www.technologyreview.com/view/525966/the-anatomy-of-a-forgotten-social-network/}

\end{abstract}

\section{Introduction}

Tumblr, as one of the most prevalent microblogging sites, has become
phenomenal in recent years, and it is acquired by Yahoo! in 2013.
By mid-January 2014, Tumblr has 166.4 millions of users and 73.4 billions of
posts\footnote{http://www.tumblr.com/about}.  It
is reported to be the most popular social site among young generation, as half of Tumblr's visitor
are under 25 years old\footnote{http://www.webcitation.org/64UXrbl8H}. Tumblr is ranked
as the 16th most popular sites in United States, which is the 2nd most
dominant blogging site, the 2nd largest microblogging service, and the
5th most prevalent social
site\footnote{http://www.alexa.com/topsites/countries/US}.  In
contrast to the momentum Tumblr gained in recent press,
little academic research
has been conducted over this burgeoning social service.   Naturally questions
arise:   \emph{What is Tumblr? What is the difference between Tumblr and other
blogging or social media sites?}

Traditional blogging sites, such as
Blogspot\footnote{http://blogspot.com} and LivingSocial\footnote{http://livesocial.com},
have high quality content but
little social interactions. Nardi
\textit{et al.}~\cite{blog2004} investigated  blogging as a form
of personal communication and expression,  and showed that the vast majority of blog posts
are written by ordinary people with a small audience.
On the contrary, popular
social networking sites like
Facebook\footnote{http://facebook.com},
have richer social interactions, but lower quality content comparing
with blogosphere. Since most social interactions are
either unpublished or less meaningful for the majority  of public
audience,  it is natural for Facebook users to form different
communities or social circles.  Microblogging services,
in between of traditional blogging and online social
networking services, have intermediate quality content and
intermediate social
interactions. Twitter\footnote{http://twitter.com}, which is the
largest microblogging site, has the limitation of 140 characters in each
post, and the Twitter following relationship is not reciprocal: a
Twitter user does not need to follow back if the user is followed by
another. As a result, Twitter is considered as a new social
media~\cite{KwakWWW2010}, and short messages can be broadcasted to a
Twitter user's followers in real time.

Tumblr is also posed as a microblogging platform.  Tumblr users can
follow another user without following back, which forms a non-reciprocal social network; a
Tumblr post can be re-broadcasted by a user to its own followers via
{\it reblogging}. But unlike Twitter, Tumblr has no length limitation for each post, and Tumblr
also supports multimedia post, such as images, audios or videos. With
these differences in mind,  are the social network, user generated content, or
user behavior on Tumblr dramatically different from other social media sites?

In this paper, we provide a statistical overview over Tumblr from
assorted aspects.  We study
the social network structure among Tumblr users and compare its
network properties with other commonly used ones.  Meanwhile, we study
content generated in Tumblr and examine the content generation patterns.
One step further, we also analyze how a blog post is being reblogged
and propagated through a network, both topologically and temporally.
Our study shows that Tumblr
provides hybrid microblogging services: it contains dual
characteristics of both social media and traditional blogging.
Meanwhile, surprising patterns surface.
We describe these intriguing
findings and provide insights, which hopefully can be leveraged by other researchers to understand
more about this new form of social media. 




\section{Tumblr at First Sight}
Tumblr is ranked the second largest microblogging service, right after
Twitter, with over 166.4 million users and 73.4 billion posts by January 2014.
Tumblr is easy to register, and one can sign up for Tumblr service with a valid email address within 30
seconds.  Once sign in Tumblr, a user can follow other
users. Different from Facebook, the connections in Tumblr do not require
mutual confirmation. Hence the social network in Tumblr is
unidirectional.




Both Twitter and Tumblr are considered as microblogging platforms. Comparing with Twitter, Tumblr exposes several differences:
\begin{itemize}
\item There is no length limitation for each post;
\item Tumblr supports multimedia
posts, such as images, audios and videos;
\item Similar to hashtags in Twitter, bloggers can also tag their blog post, which is commonplace
in traditional blogging.  But tags in Tumblr are seperate
from blog content, while in Twitter  the hashtag can appear anywhere
within a tweet.
\item Tumblr recently (Jan. 2014) allowed
users to mention and link to specific users inside posts. This {\it @user}
mechanism needs more time to be adopted by the community;
\item Tumblr does not
differentiate  verified account. 
\end{itemize}

\begin{figure}[h]
  \centering
  \includegraphics[width=0.48\textwidth]{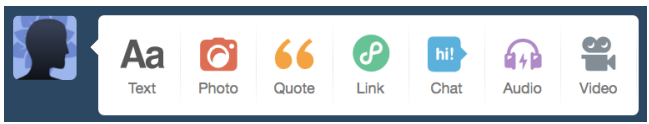}
  \caption{Post Types in Tumblr}
  \label{fig:tumblr-post-type}
\end{figure}
Specifically, Tumblr defines 8 types of posts: {\it photo}, {\it
  text}, {\it quote}, {\it audio}, {\it video}, {\it chat}, {\it link}
and {\it answer}.  As shown in Figure~\ref{fig:tumblr-post-type}, one
has the flexibility to start a post in any type except {\it
  answer}.  {\it Text}, {\it photo}, {\it audio}, {\it video} and {\it
  link} allow one to post, share and comment any multimedia content.
{\it Quote} and {\it chat}, which are not available in most other social
networking platforms, let Tumblr users share quote or chat history from
ichat or msn.  {\it Answer} occurs only when one tries to interact with
other users: when one user posts a question, in particular,
writes a post with text box ending with a question mark, the user
can enable the option for others to answer the question, which will be
disabled automatically after 7 days. A post can also be reblogged by another user to broadcast
to his own followers.  The reblogged post will quote the original
post by default and allow the reblogger to add additional comments.

\begin{figure}
  \begin{center}
    \includegraphics[width=3.0in,angle=0]{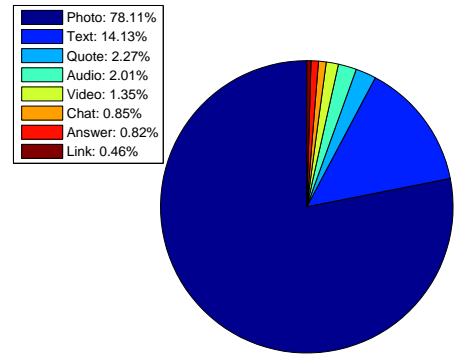}
  \end{center}
  \vspace{-0.9cm}
  \caption{Distribution of  Posts (Better viewed in color)}
  \label{fig:post_dis}
\end{figure}

Figure \ref{fig:post_dis} demonstrates the distribution of Tumblr post
types, based on 586.4 million
posts we collected.
As seen in the figure, even though all
kinds of content are supported, {\it photo} and {\it text} dominate
the distribution, accounting for more than $92\%$ of the posts. Therefore,
we will concentrate on these two types of posts for our content
analysis later.

Since Tumblr has a strong presence of photos, it is natural to compare
it to other photo or image based social networks like Flickr\footnote{http://flickr.com} and
Pinterest\footnote{http://pinterest.com}. Flickr is mainly an image hosting website, and Flicker users
can add contact, comment or like others' photos.  Yet, different from
Tumblr, one cannot reblog another's photo in Flickr.  Pinterest is
designed for curators, allowing one to share photos or videos of her
taste with the public. Pinterest links a pin to the commercial website
where the product presented in the pin can be purchased, which accounts
for a stronger e-commerce behavior. Therefore, the target audience of Tumblr and
Pinterest are quite different: the majority of users in Tumblr
are under age 25, while Pinterest is heavily used by women within age
from 25 to 44~\cite{mittal2013pin}.

We directly sample a sub-graph snapshot of social network from Tumblr on
August 2013, which contains 62.8 million
nodes and 3.1 billion edges.  Though this graph is not yet up-to-date,
we believe that many network properties should be well preserved given the
scale of this graph. Meanwhile, we sample about 586.4 million of Tumblr posts
from August 10 to September 6, 2013. Unfortunately, Tumblr does not require users to
fill in basic profile information, such as gender or location.
Therefore, it is impossible for us to conduct user profile analysis as done in
other works.
In order to handle such large volume of data, most statistical patterns are
computed through a MapReduce cluster, with some
algorithms being tricky.  We will skip the involved
implementation details but concentrate solely on the derived patterns.

\begin{figure*}
  \centering
  \subfigure[in/out degree distribution]{ \includegraphics[width=0.30\textwidth]{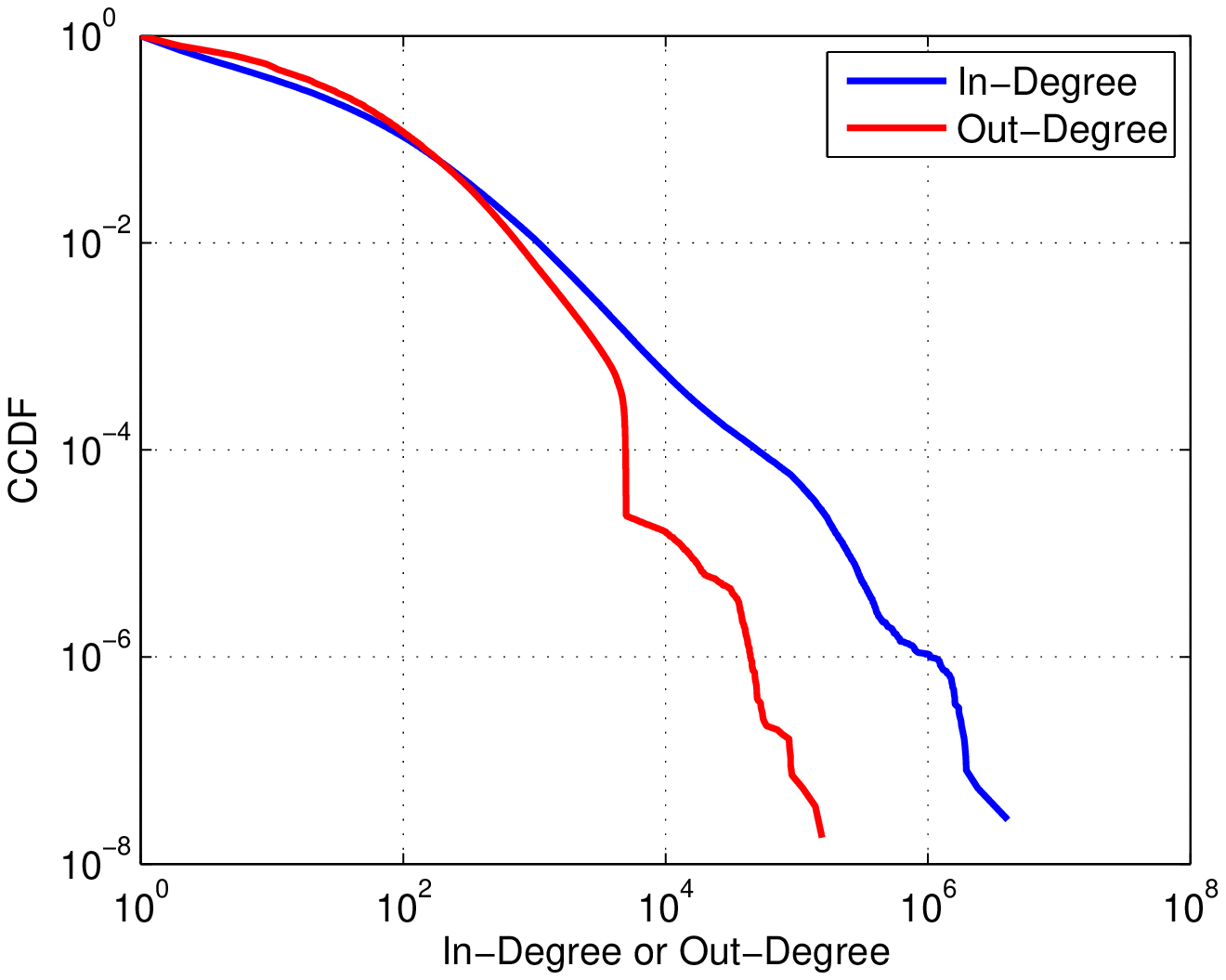} \label{fig:in_out_degree} }
  \subfigure[in/out degree correlation]{  \includegraphics[width=0.35\textwidth]{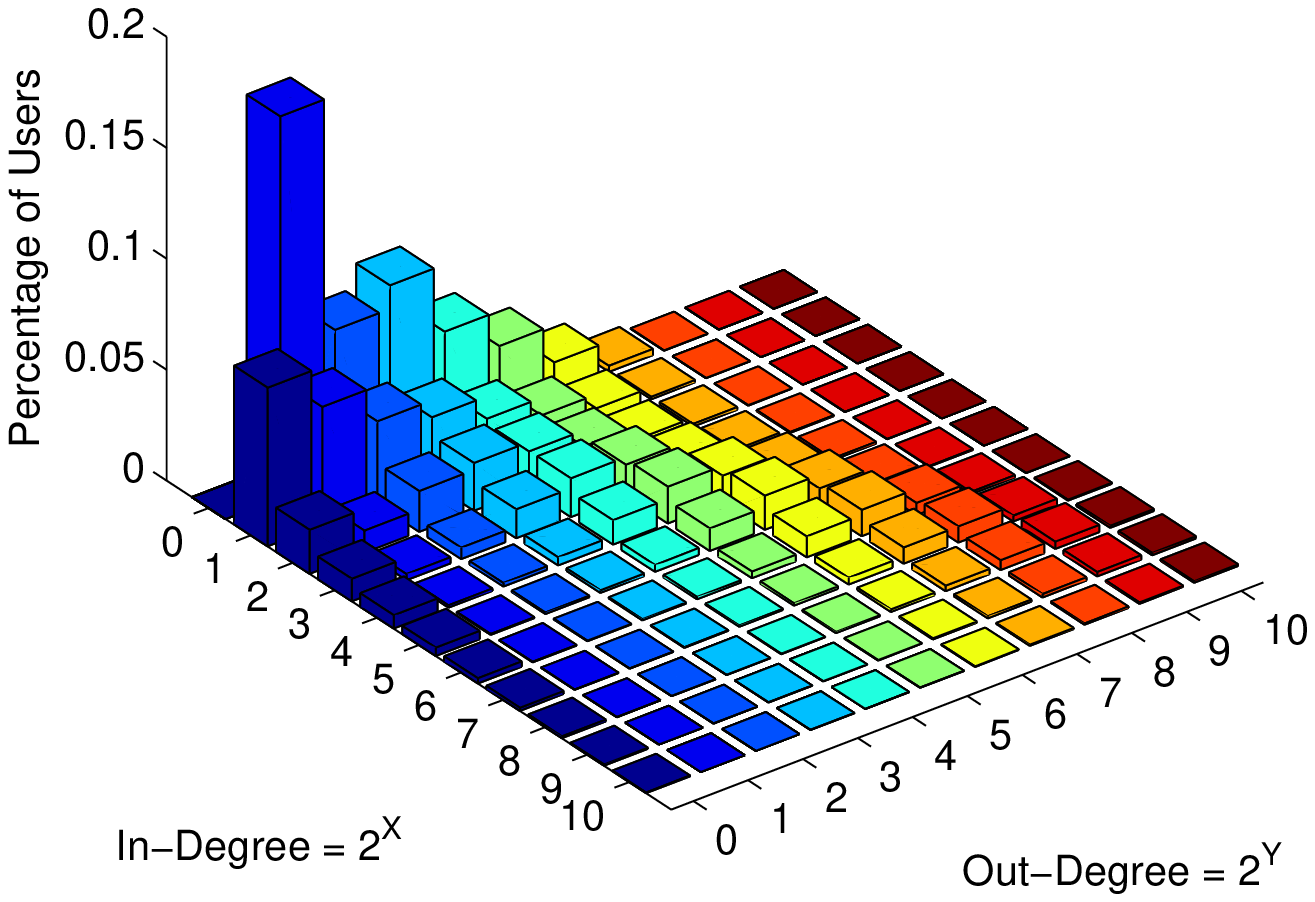} \label{fig:3d_bar} }
   \subfigure[degree distribution in r-graph]{ \includegraphics[width=0.30\textwidth]{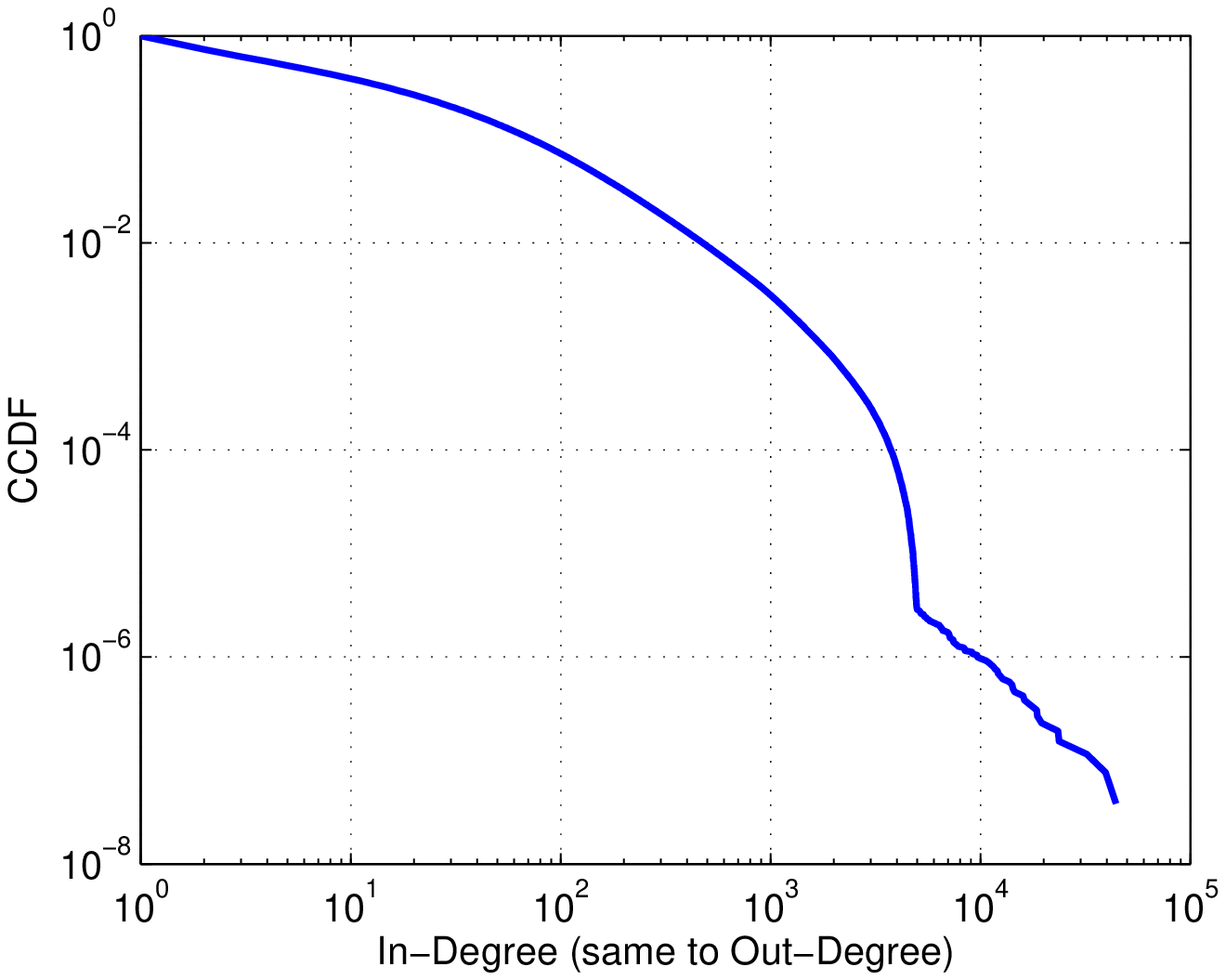} \label{fig:mutual_degree}}
   \caption{Degree Distribution of Tumblr Network}
  \label{fig:degree-distr}
\end{figure*}

Most statistical patterns can be presented in three
different forms: \textit{probability density function} (PDF),
\textit{cumulative distribution function} (CDF)  or \textit{complementary
cumulative distribution function} (CCDF),  describing $Pr(X=x)$,
$Pr(X \le x)$ and $Pr( X \ge x)$ respectively, where $X$ is a random
variable and $x$ is certain value.  Due to the space limit, it
is impossible to include all of them.  Hence, we decide which form(s) to include
depending on presentation and comparison convenience with other
relevant papers.  That is, if CCDF is reported in a relevant paper,
we try to also report CCDF here so that rigorous comparison is
possible.

Next,  we study properties of Tumblr through different lenses,
in particular, as a social network, a content generation website, and
an information propagation platform, respectively.

\section{Tumblr as Social Network}
\label{sec:social-network}
We begin our analysis of Tumblr by examining its social network
topology structure. Numerous social networks have been analyzed in
the past, such as traditional blogosphere~\cite{ShiICWSM2007},
Twitter~\cite{Java2007,KwakWWW2010},
Facebook~\cite{ugander2011anatomy}, and instant messenger
communication network~\cite{Lesk-Horv08}.
Here we run an array of standard network analysis to compare with
other networks, with results summarized in
Table~\ref{tab:network-stats}\footnote{
Even though we wish to include results over other popular social media
networks like Pinterest, Sina Weibo and Instagram,  analysis over those
websites not available or just small-scale case studies that are
difficult to generalize to a comprehensive scale for a fair
comparison. Actually in the Table,  we
  observe quite a discrepancy between numbers reported over a small
  twitter data set and another comprehensive snapshot.}.

\begin{table*}
  \centering
  \caption{Comparison of Tumblr  with other popular social
    networks. The numbers of Blogosphere, Twitter-small,
    Twitter-huge, Facebook, and MSN are obtained from
    \cite{ShiICWSM2007,Java2007,KwakWWW2010,ugander2011anatomy,Lesk-Horv08},
    respectively. In the table, -- implies the corresponding
    statistic is not available or not applicable; GCC denotes the giant connected
    component;  the symbols in parenthesis \emph{m, d, e, r } respectively represent \emph{mean}, \emph{median}, \emph{the $90\%$ effective diameter}, and  \emph{diameter} (the maximum shortest path in the network).
}
  \label{tab:network-stats}
  \begin{tabular}{c|rrrrrr}
    \hline
    Metric & Tumblr & Blogosphere & Twitter-small & Twitter-huge &
    Facebook & MSN \\
    \hline
\#nodes & 62.8M & 143,736  &   87,897 & 41.7M  & 721M & 180M \\
\#links   & 3.1B    &  707,761 & 829,467 &  1.47B  &  68.7B & 1.3B \\
in-degree distr &  $\propto k^{-2.19}$  &  $\propto k^{-2.38}$
& $\propto k^{-2.4}$ & $\propto k^{-2.276}$ &  -- &  -- \\
degree distr in r-graph & $\ne$ power-law & -- & -- & -- & $\ne$ power-law &$\propto k^{0.8} e^{-0.03k}$ \\
\hline
direction &  directed & directed & directed & directed & undirected & undirected \\
reciprocity   &  29.03\% & 3\% & 58\% &  22.1\% & -- &  -- \\
degree correlation &  0.106 &  -- & -- & $>0$ & 0.226 & --  \\
\hline
avg distance & 4.7(m), 5(d) & 9.3(m) & -- & 4.1(m), 4(d) & 4.7(m), 5(d) & 6.6(m), 6(d) \\
diameter & 5.4(e), $\ge 29$(r) & 12(r) & 6(r) & 4.8(e), $\ge 18$(r)&  $<5$(e) & 7.8(e), $\ge 29$(r) \\
GCC coverage & 99.61\% & 75.08\% & 93.03\% &  -- & 99.91\% & 99.90\%\\
\hline
  \end{tabular}
\end{table*}




{\bf Degree Distribution.} Since Tumblr does not require mutual confirmation when one follows
another user, we represent the follower-followee network in Tumblr as
a directed graph: in-degree of a user represents how many
followers the user has attracted, while out-degree indicates how
many other users one user has been following.  Our sampled sub-graph contains
62.8 million nodes and 3.1 billion edges. Within this social graph, 41.40\%
of nodes have 0 in-degree, and the maximum in-degree of a node is 4.06
million. By contrast,  12.74\% of nodes have 0 out-degree, the maximum out-degree of
a node is 155.5k.  Top popular Tumblr users include {\it
  equipo}\footnote{http://equipo.tumblr.com}, {\it instagram}\footnote{http://instagram.tumblr.com}, and
{\it woodendreams}\footnote{http://woodendreams.tumblr.com}.
This indicates the media characteristic of
Tumblr: the most popular user has more than 4 million audience,
while more than 40\% of users are purely audience since they
don't have any followers.


Figure \ref{fig:in_out_degree} demonstrates the distribution of
in-degrees in the blue curve and that of out-degrees in
the red curve, where y-axis refers to the cumulated density
distribution function (CCDF): the probability that
accounts have at least k in-degrees or out-degrees, i.e., $P(K>=k)$.
It is observed that Tumblr users' in-degree follows
a power-law distribution with exponent $-2.19$, which is quite
similar from the power law exponent of Twitter at $-2.28$
~\cite{KwakWWW2010} or that of traditional blogs at
$-2.38$~\cite{ShiICWSM2007}.   This also confirms with earlier empirical
observation that most social network have a power-law exponent between
$-2$ and $-3$~\cite{Clau-etal07}.

In regard to out-degree distribution,
we notice the red curve has a big drop when out-degree is around 5000, since
there was a limit that ordinary Tumblr users can follow at most 5000
other users. Tumblr users' out-degree does not follow a power-law
distribution, which is similar to blogosphere of traditional
blogging~\cite{ShiICWSM2007}.


If we explore user's in-degree and out-degree together, we could
generate normalized 3-D histogram in Figure \ref{fig:3d_bar}. As both
in-degree and out-degree follow the heavy-tail distribution, we only
zoom in those user who have less than $2^{10}$ in-degrees and
out-degrees.  Apparently, there is a positive correlation between in-degree and
out-degree because of the dominance of diagonal bars.  In aggregation,
a user with low in-degree tends to have low out-degree as
well, even though some nodes, especially those top popular ones,  have very imbalanced in-degree and
out-degree.


{\bf Reciprocity.}
Since Tumblr is a directed network, we would like to
examine the reciprocity of the graph.  We derive the backbone of  the
Tumblr network by keeping those reciprocal connections only, i.e., user $a$
follows $b$ and vice versa. Let \textit{r-graph}
denote the corresponding reciprocal graph.   We found
29.03\% of Tumblr user pairs have
reciprocity relationship, which is higher than 22.1\% of reciprocity
on Twitter~\cite{KwakWWW2010} and 3\% of reciprocity on
Blogosphere~\cite{ShiICWSM2007}, indicating a stronger interaction
between users in the network.
Figure \ref{fig:mutual_degree} shows
the distribution of degrees in the r-graph. There is a turning point
due to the Tumblr limit of $5000$ followees for ordinary users. The reciprocity relationship
on Tumblr does not follow the power law distribution, since the curve
mostly is convex, similar to the pattern reported over
Facebook\cite{ugander2011anatomy}.

Meanwhile,  it has been observed that one's degree is correlated with
the degree of his friends. This is also called {\it degree correlation} or
{\it degree assortativity}~\cite{newman2002assortative,newman2003mixing}.  Over the derived r-graph, we obtain a
correlation of $0.106$ between terminal nodes of reciprocate connections,
reconfirming the positive degree assortativity as reported in Twitter~\cite{KwakWWW2010}.  Nevertheless, compared with
the strong social network Facebook, Tumblr's degree assortativity is
weaker~($0.106$ vs. $0.226$).

\begin{figure}[h]
  \centering
  \includegraphics[width=0.3\textwidth]{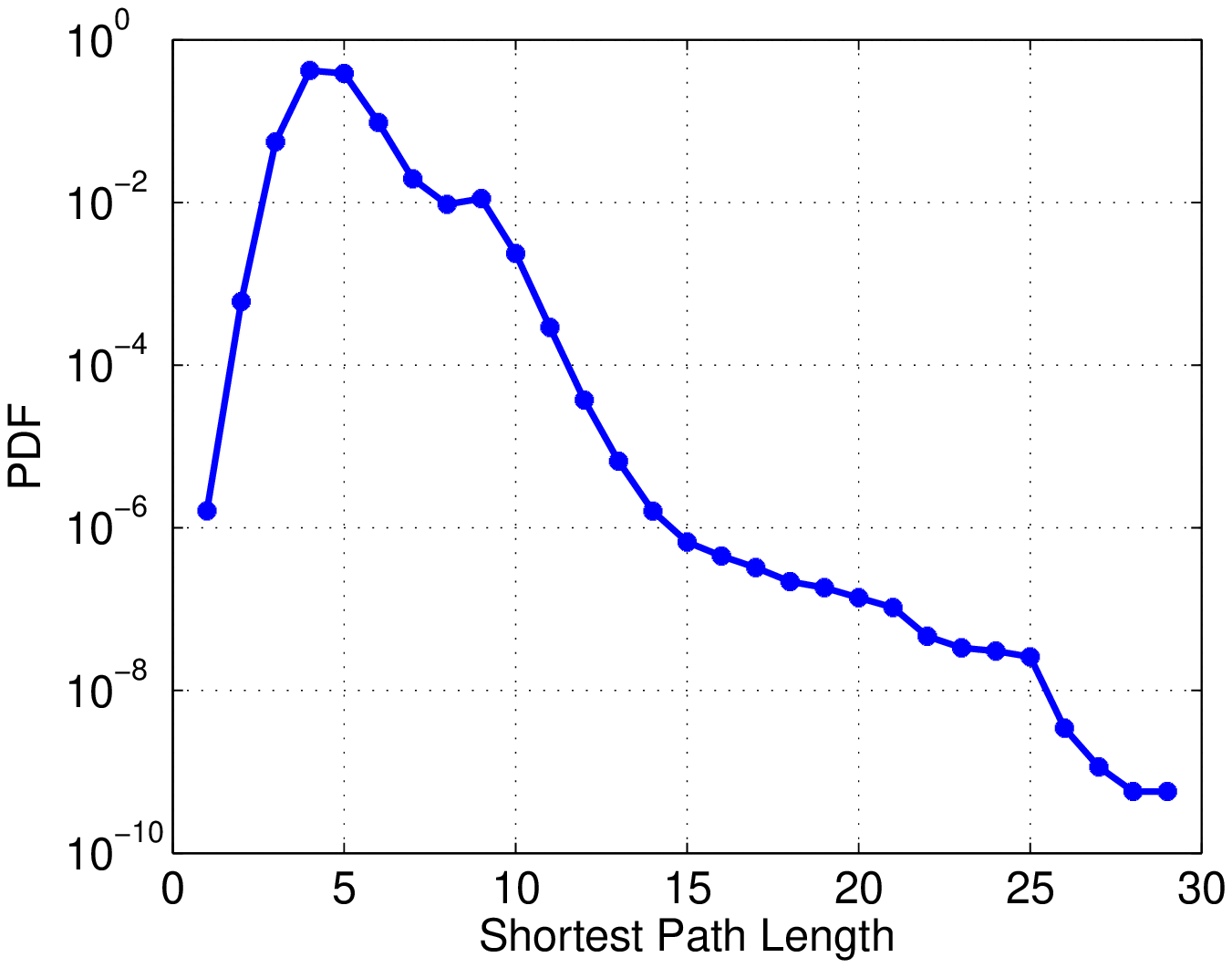}
  \includegraphics[width=0.3\textwidth]{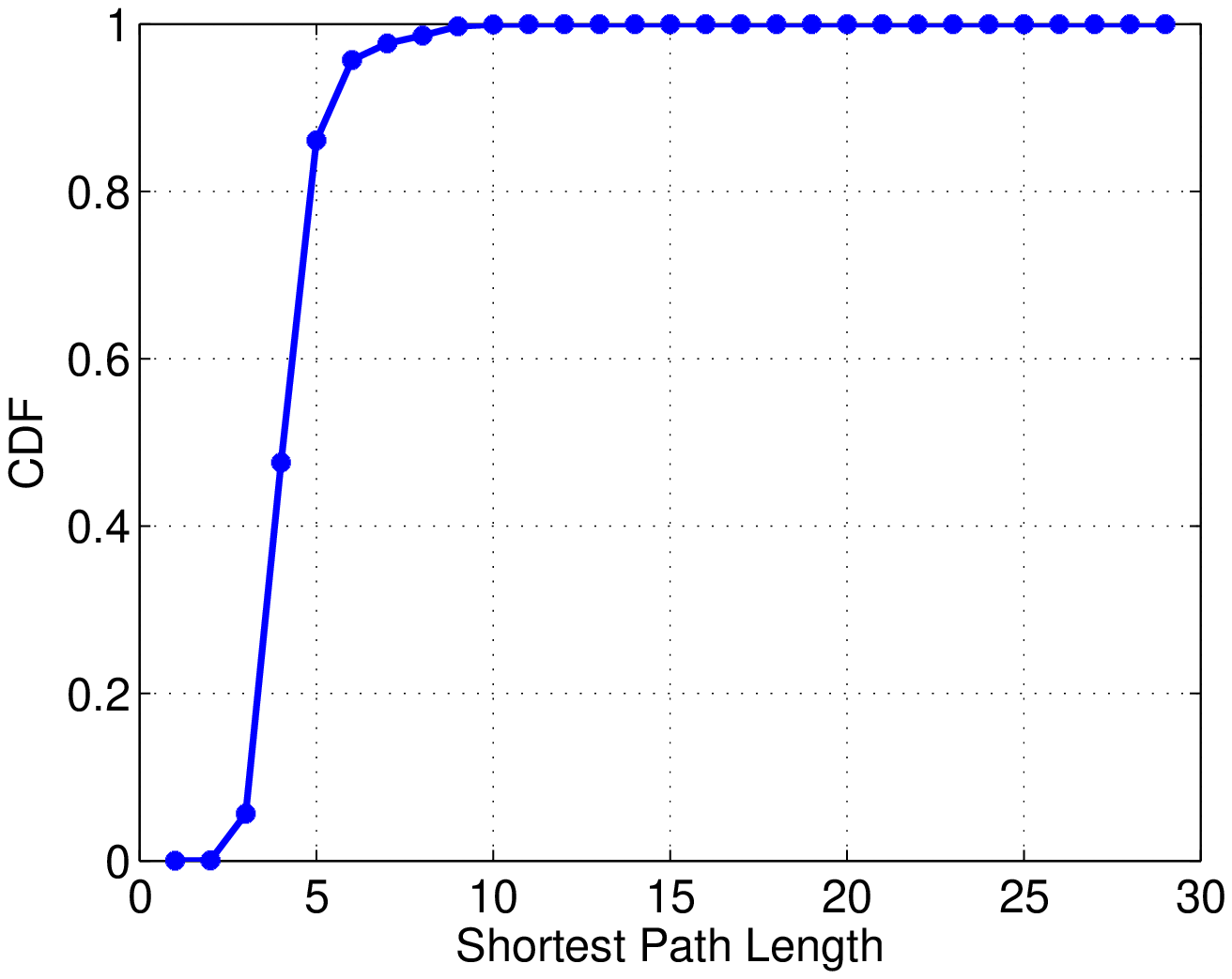}
  \caption{Shortest Path Distribution}
\label{fig:shortest-path}
\end{figure}

{\bf Degree of Separation.}
Small world phenomenon is almost universal among social networks. With
this huge Tumblr network, we are able to validate the well-known ``six degrees
of separation'' as well.  Figure~\ref{fig:shortest-path} displays the
distribution of the shortest paths in the network.  To approximate the
distribution, we randomly sample 60,000 nodes as seed and calculate  for
each  node the shortest paths to other nodes.  It is observed that the
distribution of paths length reaches its mode with the highest probability at $4$ hops, and has a
median of $5$ hops. On average, the distance between two connected
nodes is $4.7$.  Even though the longest shortest path in the approximation has
$29$ hops, $90\%$ of shortest paths are within $5.4$ hops.  All these
numbers are close to those reported on Facebook and Twitter, yet
significantly smaller than that obtained over blogosphere and instant
messenger network~\cite{Lesk-Horv08}.

{\bf Component Size.}  The previous result shows that those users who
are connected have a small average distance. It relies on the
assumption that most users are connected to each other, which we shall
confirm immediately.  Because the Tumblr graph is directed, we compute
out all weakly-connected components by ignoring the direction of
edges.  It turns out the giant connected component (GCC) encompasses
$99.61\%$ of nodes in the graph.  Over the derived r-graph, $97.55\%$
are residing in the corresponding GCC. This finding suggests the
whole graph is almost just one connected component, and almost all
users can reach others through just few hops.


To give a palpable understanding, we summarize commonly used network statistics in
Table~\ref{tab:network-stats}. Those numbers from other popular social
networks (blogosphere, Twitter, Facebook, and MSN) are also included
for comparison. From this compact view, it is obvious traditional
blogs yield a significantly different network structure. Tumblr, even
though originally proposed for blogging, yields a network
structure that is more similar to Twitter and Facebook.

\section{Tumblr as Blogosphere for \\Content Generation}
As Tumblr is initially proposed for the purpose of blogging, here we analyze its user generated
contents.
As described earlier, \textit{photo} and
\textit{text} posts account for more than $92\%$ of total posts.
Hence, we concentrate only on these two types of posts.  One text post
may contain URL, quote or raw message. In this study, we are mainly
interested in the authentic contents generated by users. Hence, we extract raw
messages as the content information of each text post, by removing quotes and URLs.  Similarly,
photo posts contains 3 categories of information: photo URL, quote
photo caption, raw photo caption. While the photo URL might contain
lots of additional meta information, it would require tremendous effort
to analyze all images in Tumblr.  Hence, we focus on raw photo captions as the content of each photo post.
We end up with two datasets of content: one is \textit{text post}, and the other is \textit{photo caption}.


\begin{table}
  \centering
  \begin{tabular}{|c|c|c|}
    \hline
    & Text Post  & Photo Caption  \\
    & Dataset & Dataset \\
    \hline
    \# Posts & 21.5 M & 26.3 M \\
    \hline
    Mean Post Length & 426.7 Bytes & 64.3 Bytes \\
    \hline
    Median Post Length & 87 Bytes & 29 Bytes \\
    \hline
    Max Post Length & 446.0 K Bytes & 485.5 K Bytes\\
    \hline
  \end{tabular}
  \caption{Statistics of User Generated Contents}
  \label{tab:stat_dataset}
\end{table}

{\bf What's the effect of no length limit for post?}   Both Tumblr and
Twitter are considered microblogging platforms, yet there is one key
difference:  Tumblr has no length limit while Twitter enforces the
strict limitation of  140 bytes for each tweet. How does this key difference affect user post behavior?

It has been reported
that the average length of posts on Twitter is 67.9 bytes and the median is 60 bytes\footnote{http://www.quora.com/Twitter-1/What-is-the-average-length-of-a-tweet}.
Corresponding statistics of Tumblr are shown in
Table~\ref{tab:stat_dataset}.  For the text post dataset,  the average
length is 426.7 bytes and the median is 87 bytes, which both, as expected,
are  longer than
that of Twitter.  Keep in mind Tumblr's numbers are obtained after removing all quotes,
photos and URLs, which further discounts the discrepancy between
Tumblr and Twitter. The big gap between mean and median is due to a small percentage of
extremely long posts. For instance, the longest text post is
446K bytes in our
sampled dataset.  As for photo captions, naturally we expect it to be much
shorter than text posts. The average length is around 64.3
bytes, but the median is only 29 bytes. Although photo posts are
dominant in Tumblr,
the number of text posts and photo captions in Table~\ref{tab:stat_dataset}
are comparable, because majority of photo posts don't contain any raw photo captions.

\begin{figure}
  \centering
  \includegraphics[width=0.3\textwidth]{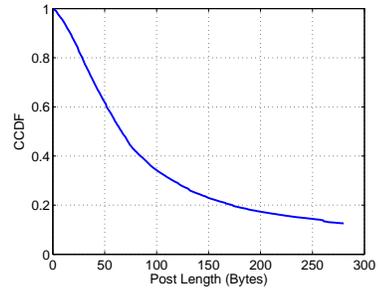}
\caption{Post Length Distribution}
  \label{fig:post-length}
\end{figure}

A further related question: \textit{is the 140-byte limit sensible?}    We plot
post length distribution of the text post dataset, and zoom into less than 280 bytes in
Figure~\ref{fig:post-length}.  
About $24.48\%$ of posts are beyond 140 bytes,
which indicates that at least around one
quarter of posts will have to be rewritten in a more compact version
if the limit was enforced in Tumblr.

Blending all numbers above together, we can see at least two types of
posts: one is more like posting a reference (URL or photo) with added
information or short comments, the other is authentic user generated
content like in traditional blogging. In other words, Tumblr is a mix of both types of posts,
and its no-length-limit policy encourages its users to post
longer high-quality content directly.

{\bf What are people talking about?}  Because there is no length limit
on Tumblr, the blog post tends to be more meaningful, which allows us
to run topic analysis over the two datasets to have an overview of the
content. We run LDA~\cite{LDA2003} with 100 topics on both datasets, and showcase several
topics and their corresponding keywords on Tables~\ref{tab:text_LDA_key} and \ref{tab:photo_LDA_key}, which also show the high quality of textual content on Tumblr clearly. \textit{Medical, Pets,
  Pop Music, Sports} are shared interests across 2 different datasets,
although representative topical keywords might be different even for
the same topic. \textit{Finance, Internet} only attracts enough
attentions from text posts, while only significant amount of photo
posts show interest to \textit{Photography, Scenery} topics. We
want to emphasize that most of these keywords are semantically
meaningful and representative of the topics.

\begin{table}[t]
  \begin{tabular}{|l | l|}
    \hline
    Topic & Topical  Keywords \\
    \hline
    Pop  & music song listen iframe  band album lyrics\\
    Music&  video guitar   \\
    \hline
    Sports & game play team win video cookie \\
    & ball football top sims fun beat league \\
    \hline
    Internet & internet computer laptop google search online \\
    & site facebook drop website app mobile iphone  \\
    \hline
    Pets & big dog cat animal pet animals bear tiny \\
    & small deal puppy \\
    \hline
    Medical & anxiety pain hospital mental panic cancer \\
    & depression brain stress medical  \\
    \hline
    Finance & money pay store loan online interest buying \\
    & bank apply card credit \\
    \hline
  \end{tabular}
  \caption{Topical Keywords from Text Post Dataset}
  \label{tab:text_LDA_key}
\end{table}

\begin{table}[t]
  \begin{tabular}{|l | l|}
    \hline
    Topic & Topical Keywords \\
    \hline
    Pets & cat dog cute upload kitty batch puppy \\
    & pet animal kitten adorable \\
    \hline
    Scenery & summer beach sun sky sunset sea nature  \\
    & ocean island clouds lake pool beautiful  \\
    \hline
    Pop  & music song rock band album listen lyrics\\
    Music& punk guitar dj pop sound hip \\
    \hline
    Photography & photo instagram pic picture check \\
    & daily shoot tbt photography \\
    \hline
    Sports & team world ball win football club  \\
    & round false soccer league baseball \\
    \hline
    Medical & body pain skin brain depression hospital \\
    & teeth drugs problems sick cancer blood \\
    \hline
  \end{tabular}
  \caption{Topical Keywords from Photo Caption Dataset}
  \label{tab:photo_LDA_key}
\end{table}

{\bf Who are the major contributors of contents?} There are two potential
hypotheses. 1) One supposes those {\it socially popular users} post more.  This is derived from
the result that those popular users are followed by many users, therefore blogging is one
way to attract more audience as followers.  Meanwhile, it might be true
that blogging is an incentive for celebrities to interact or reward their
followers. 2) The other assumes that {\it long-term users} (in terms of
registration time)
post more, since they are accustomed to this service, and they are
more likely to have their own focused communities or social
circles. These peer interactions encourage them to generate more
authentic content to share with others.

\begin{figure}[t]
  \centering
  \includegraphics[width=0.45\textwidth]{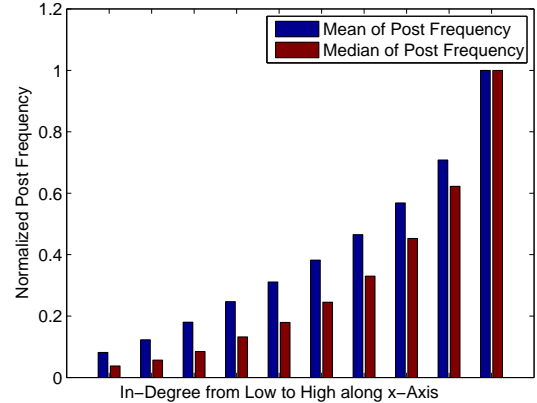} \label{fig:equidepth_postfreq_indegree}
  \includegraphics[width=0.45\textwidth]{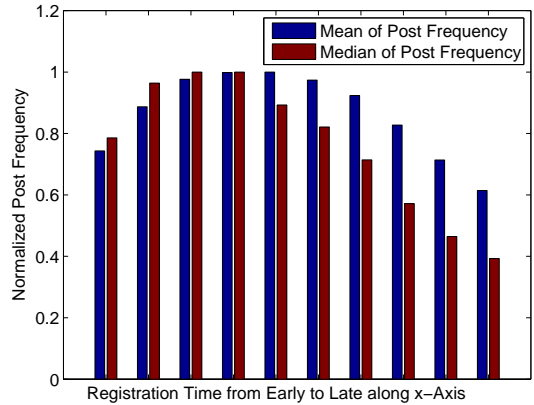} \label{fig:equidepth_postfreq_time}
  \caption{Correlation of Post Frequency with User In-degree or Duration Time
    since Registration}
  \label{fig:post_freq}
\end{figure}

Do socially popular users or long-term users generate more contents?
In order to answer this question, we choose a fixed time window of two weeks in
August 2013 and examine how frequent each user blogs on Tumblr.  We
sort all users based on their in-degree (or duration time since registration) and
then partition them  into 10 equi-width bins.  For each bin, we
calculate the average blogging frequency.  For easy comparison, we
consider the maximal value of all bins as 1, and normalize the relative
ratio for other bins.  The results are displayed
in Figure~\ref{fig:post_freq}, where x-axis from left to right
indicates increasing in-degree (or decreasing duration time). For brevity, we
just show the result for text post dataset as similar patterns were
observed over photo captions.

The patterns are strong in both figures. Those users who have
higher in-degree tend to post more, in terms of both mean and
median. One caveat is that what we observe and report here is merely
correlation, and it does not derive causality.  Here we draw a
conservative conclusion that the social popularity is highly positively correlated with user
blog frequency. A similar positive correlation is also observed in
Twitter\cite{KwakWWW2010}.

In contrast, the pattern in terms of user
registration time is beyond our imagination until we draw the
figure. Surprisingly, those users who either register earliest or register latest
tend to post less frequently.  Those who are in between are
inclined to post more frequently.  Obviously, our initial hypothesis about
the incentive for new users to blog more is invalid.  There could be
different explanations in hindsight. Rather than guessing
the underlying explanation, we decide to leave this phenomenon as an open
question to future researchers.

\begin{figure}[t]
  \centering
  \includegraphics[width=0.45\textwidth]{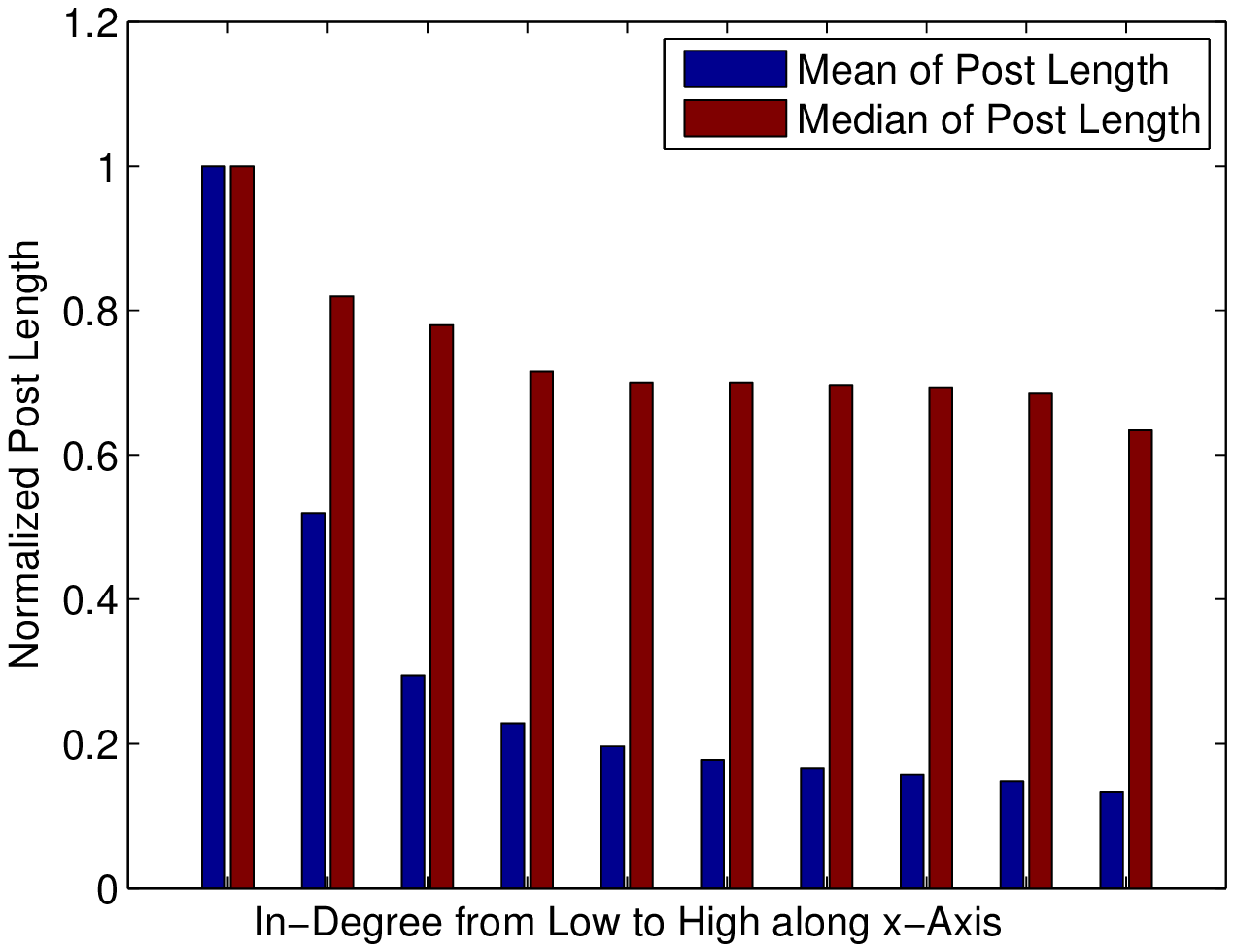} \label{fig:equidepth_postlen_indegree}
  \includegraphics[width=0.45\textwidth]{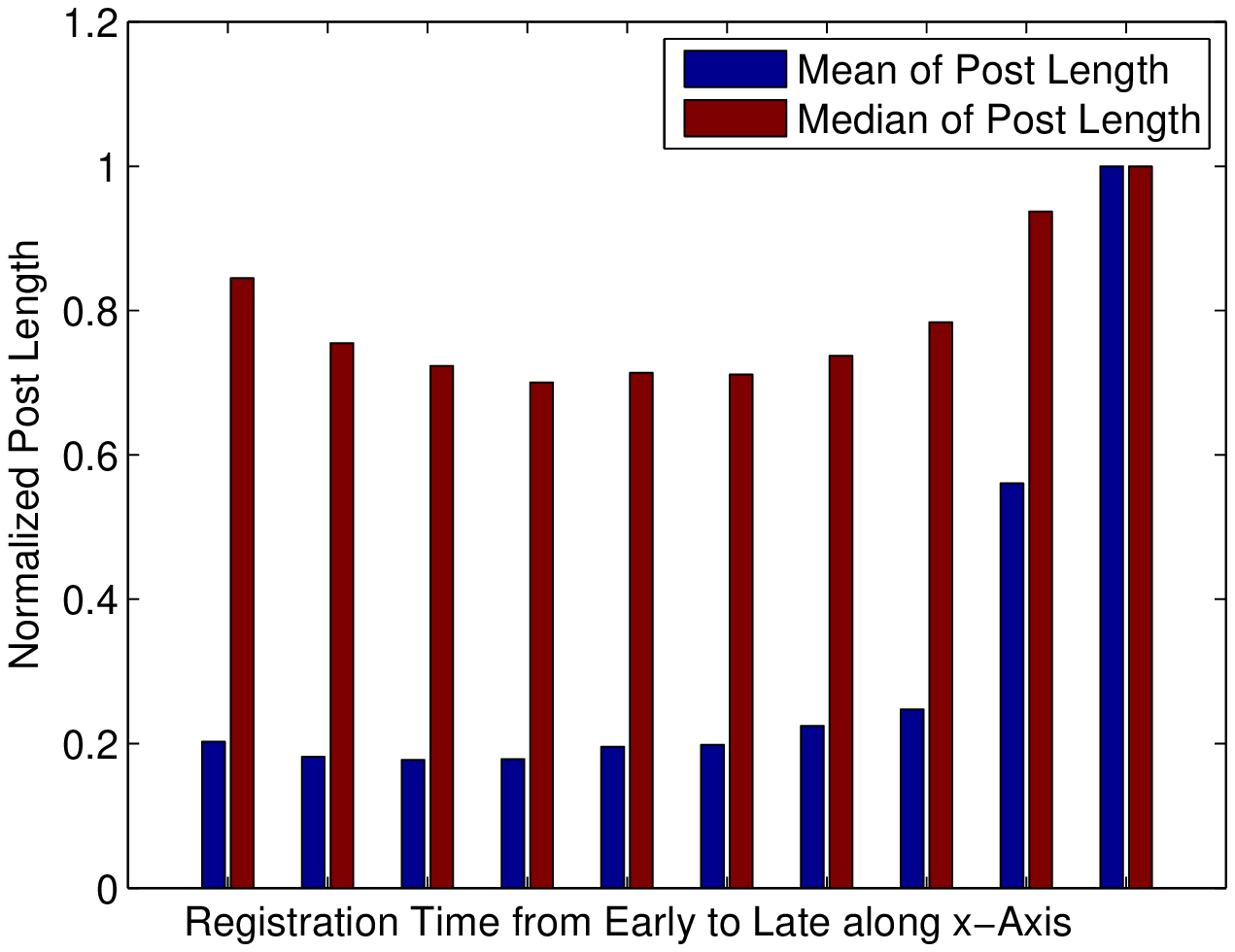} \label{fig:equidepth_postlen_time}
  \caption{Correlation of Post Length with User In-degree or Duration Time
    since Registration} \label{fig:length_indegree}
\end{figure}

As for reference, we also look at average post-length of users,
because it has been adopted as a simple metric to approximate
quality of blog posts~\cite{Agarwal:2008}.  The corresponding
correlations are plot in Figure~\ref{fig:length_indegree}.  In terms of
post length,  the tail users in social networks are the winner.
Meanwhile, long-term or recently-joined users tend to post longer
blogs. Apparently,  this pattern is exactly opposite to post
frequency. That is, the more frequent one blogs, the shorter the blog
post is. And less frequent bloggers tend to have longer posts.  That
is totally valid considering each individual has limited time and
resources.  We even changed the post length to the maximum for each
individual user rather than average, but the pattern remains still.


In summary,  without the post length limitation,  Tumblr users are
inclined to write longer blogs, and thus leading to higher-quality
user generated content, which can be leveraged for topic analysis.
The social celebrities (those with large number of followers) are the
main contributors of contents, which is similar to Twitter~\cite{DuncanWWW2011}.  Surprisingly, long-term users and
recently-registered users tend to blog less frequently.  The
post-length in general has a negative correlation with post
frequency. The more frequently one posts, the shorter those posts tend
to be.

\section{Tumblr for Information Propagation}
Tumblr offers one feature which is missing in traditional blog services:
{\it reblog}.  Once a user posts a blog, other users in Tumblr can reblog to
comment or broadcast to their own followers. This enables information to be propagated through the
network.  In this section, we examine the reblogging patterns in
Tumblr. We examine all blog posts uploaded within the first 2 weeks, and
count reblog events in the subsequent 2 weeks right
after the blog is posted, so that there would be no bias because of the time
window selection in our blog data.
%


\begin{figure}[t]
  \centering
  \includegraphics[width=0.45\textwidth]{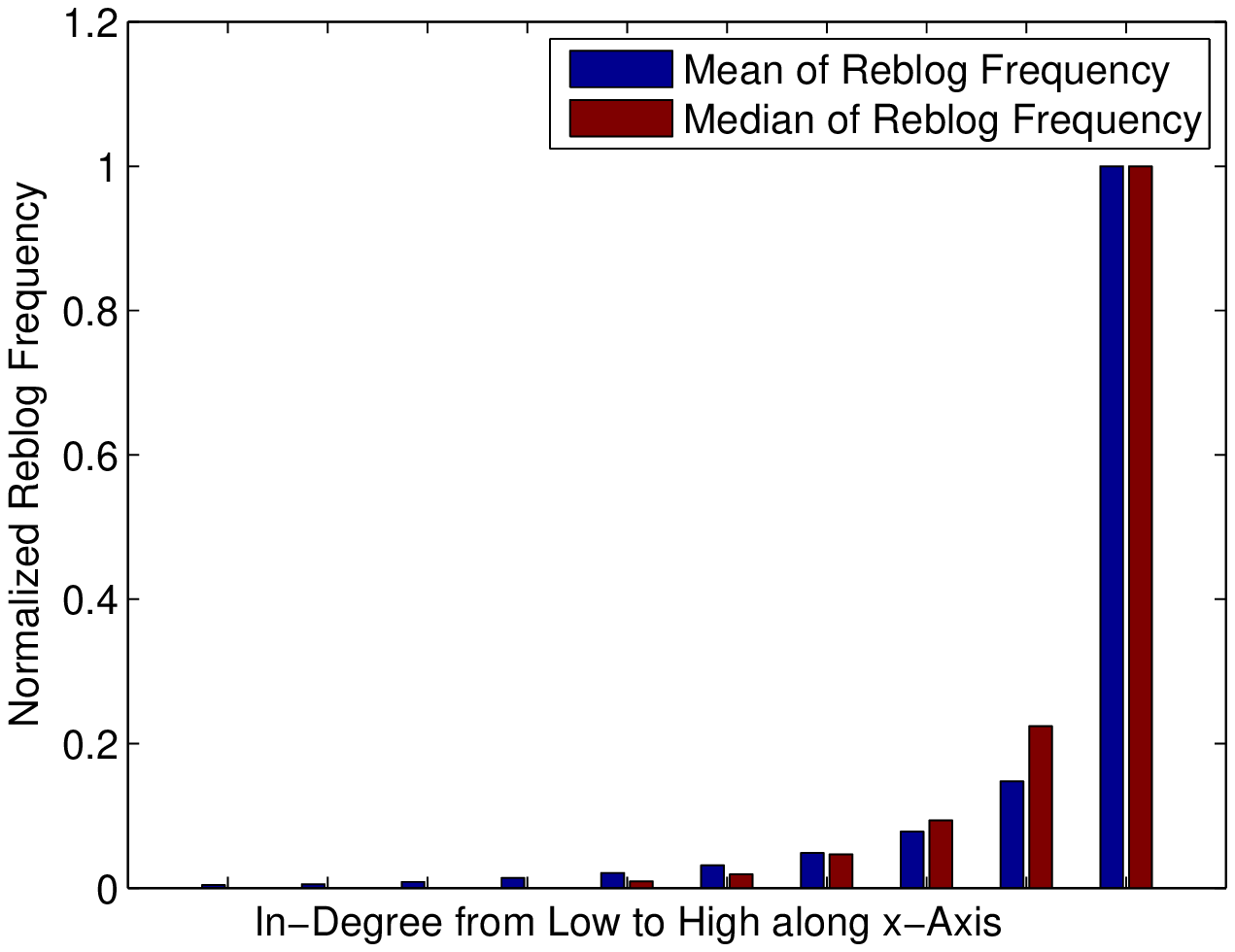}
  \includegraphics[width=0.45\textwidth]{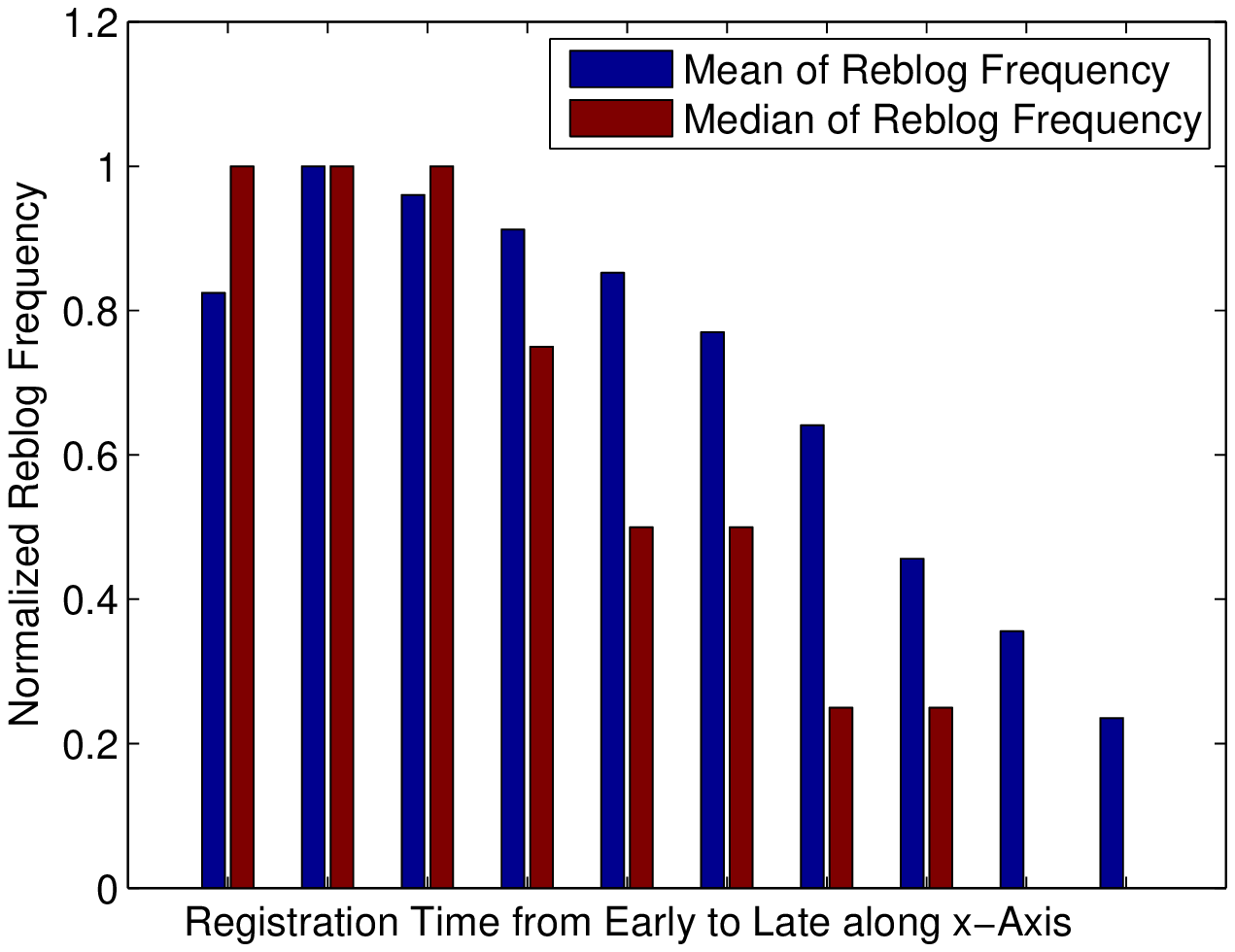} \label{fig:equidepth_reblog_time}
  \caption{Correlation of Reblog Frequency with User In-degree or Duration Time
    since Registration} \label{fig:reblog-freq}
\end{figure}

{\bf Who are reblogging?} Firstly, we would like to understand which
users tend to reblog more? Those people who reblog frequently serves
as the information transmitter. Similar to the previous section, we
examine the correlation of reblogging behavior with users'
in-degree. As shown in the Figure~\ref{fig:reblog-freq}, social
celebrities, who are the major source of contents,
reblog a lot more compared with
other users. This reblogging is propagated further through their huge
number of followers. Hence,  they serve as both content contributor
and information transmitter.
On the other hand, users who registered
earlier reblog more as well.  The socially popular and long-term users are the backbone
of Tumblr network to make it a vibrant community for information
propagation and sharing.

\begin{figure}[t]
  \centering
 \subfigure{\epsfig{file=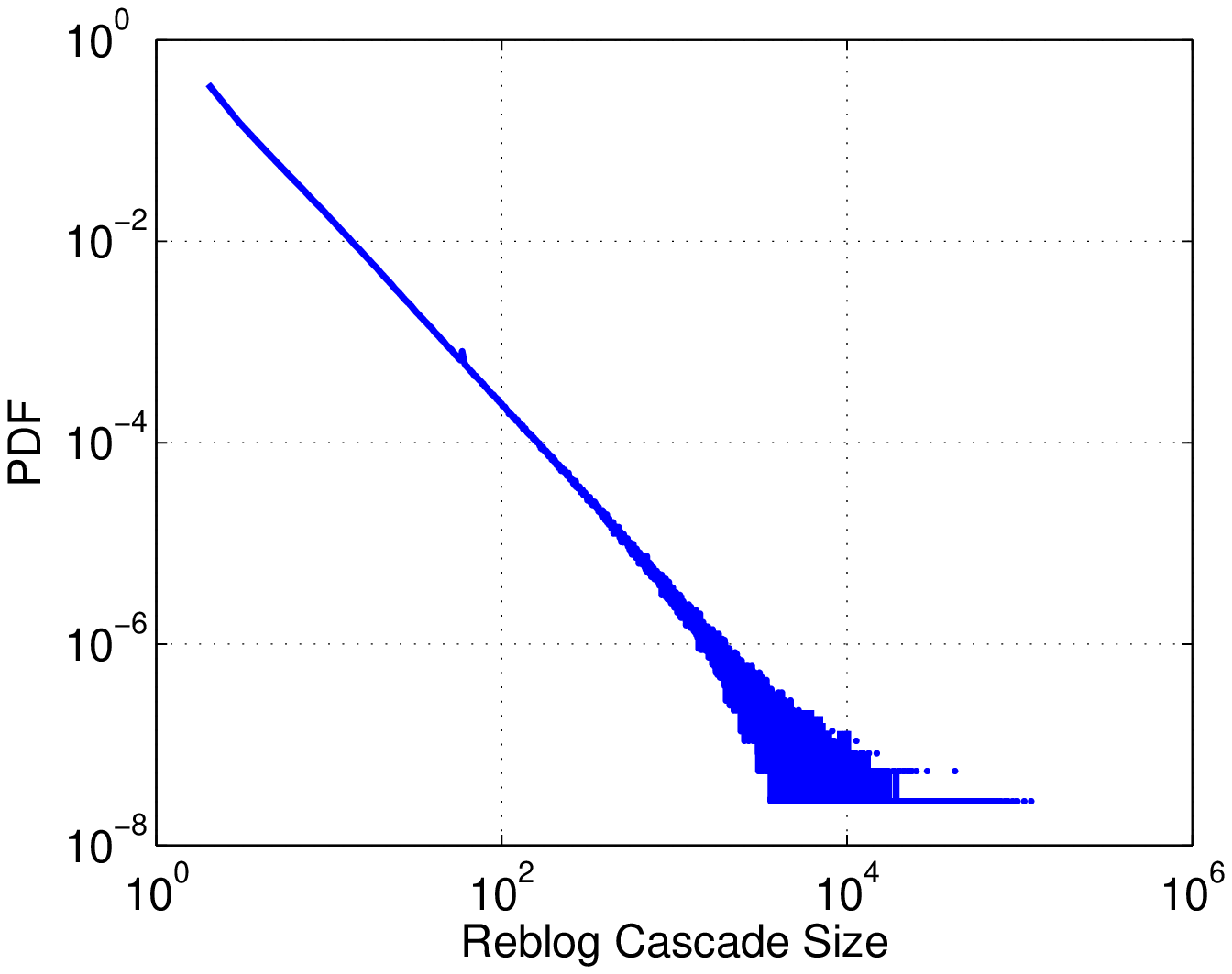,width=0.3\textwidth}
  \label{fig:tree-size-pdf}}
\subfigure{\epsfig{file=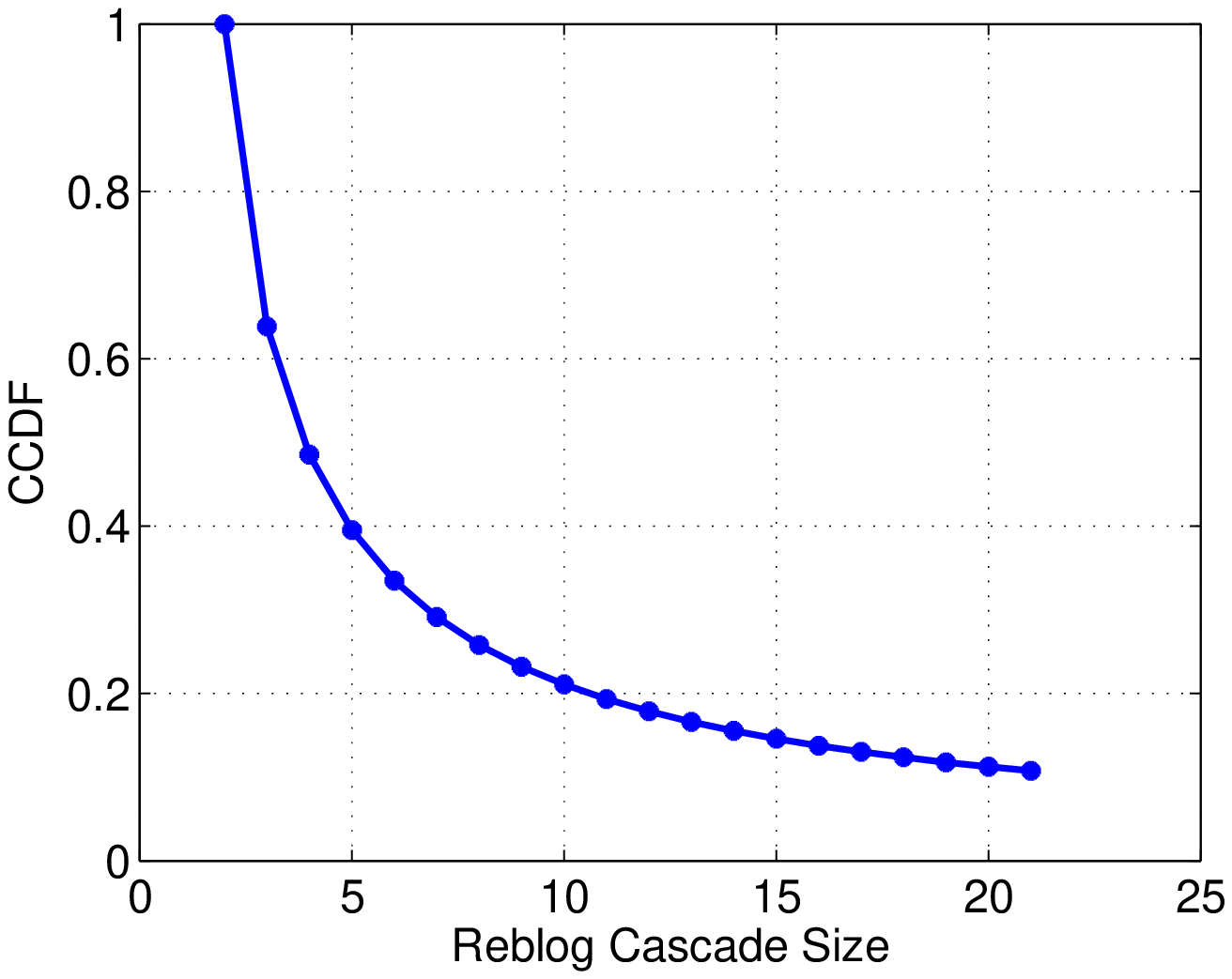,width=0.3\textwidth}
 \label{fig:tree-ccdf-top20}}
\caption{Distribution of Reblog Cascade Size}
  \label{fig:tree-size}
\end{figure}

{\bf Reblog size distribution.}  Once a blog is posted, it can be reblogged
by others. Those reblogs can be reblogged even further, which leads to
a tree structure, which is called reblog cascade, with the first author being the root
node.
The reblog cascade size indicates the number of reblog
actions that have been involved in the cascade.
Figure~\ref{fig:tree-size} plots the distribution of reblog cascade sizes.
Not surprisingly,
it follows a power-law distribution, with majority of reblog cascade
involving few reblog events.  Yet, within a time window of two weeks, the maximum cascade
could reach $116.6K$. In order to have a detailed understanding of
reblog cascades, we zoom into the short head and plot the CCDF up to
reblog cascade size equivalent  to $20$ in Figure~\ref{fig:tree-size}.  It is observed that only about $19.32\%$ of reblog
cascades have size greater than $10$.   By contrast, only $1\%$ of retweet cascades have
size larger than $10$~\cite{KwakWWW2010}.
The reblog cascades in Tumblr tend to be larger than retweet cascades in Twitter.

{\bf Reblog depth distribution.} As shown  in previous sections, almost any pair of users
are connected through few hops. How many hops does one blog to
propagate to another user in reality?  Hence, we look at the reblog cascade depth, the
maximum number of nodes to pass in order to reach one leaf node from
the root node in the reblog cascade structure.  Note that reblog depth
and size are different.  A cascade of depth $2$ can involve hundreds
of nodes if every other node in the cascade reblogs the same root node.

\begin{figure}[t]
  \centering
 \subfigure{\epsfig{file=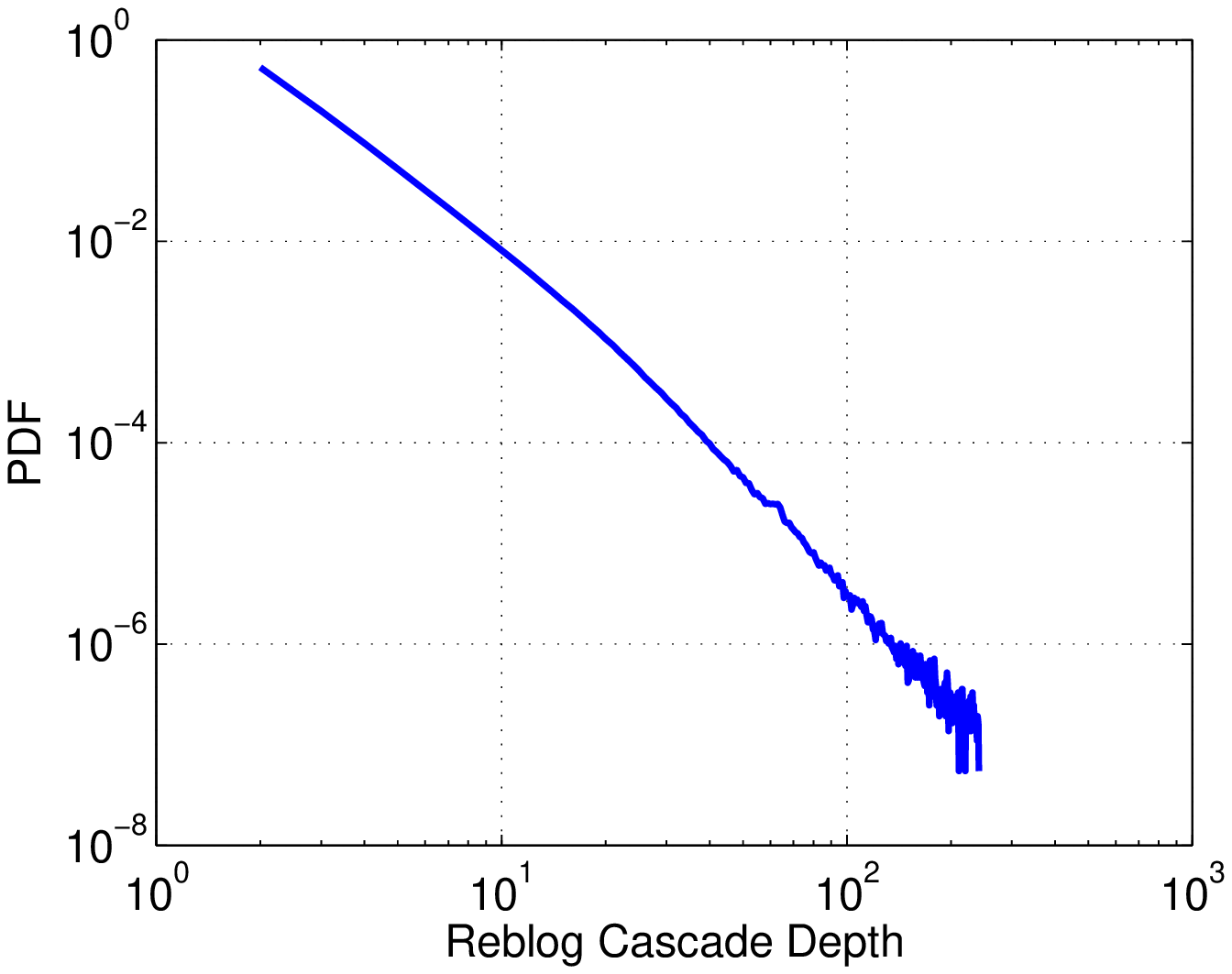,width=0.3\textwidth}
  \label{fig:tree-depth-pdf}}
\subfigure{\epsfig{file=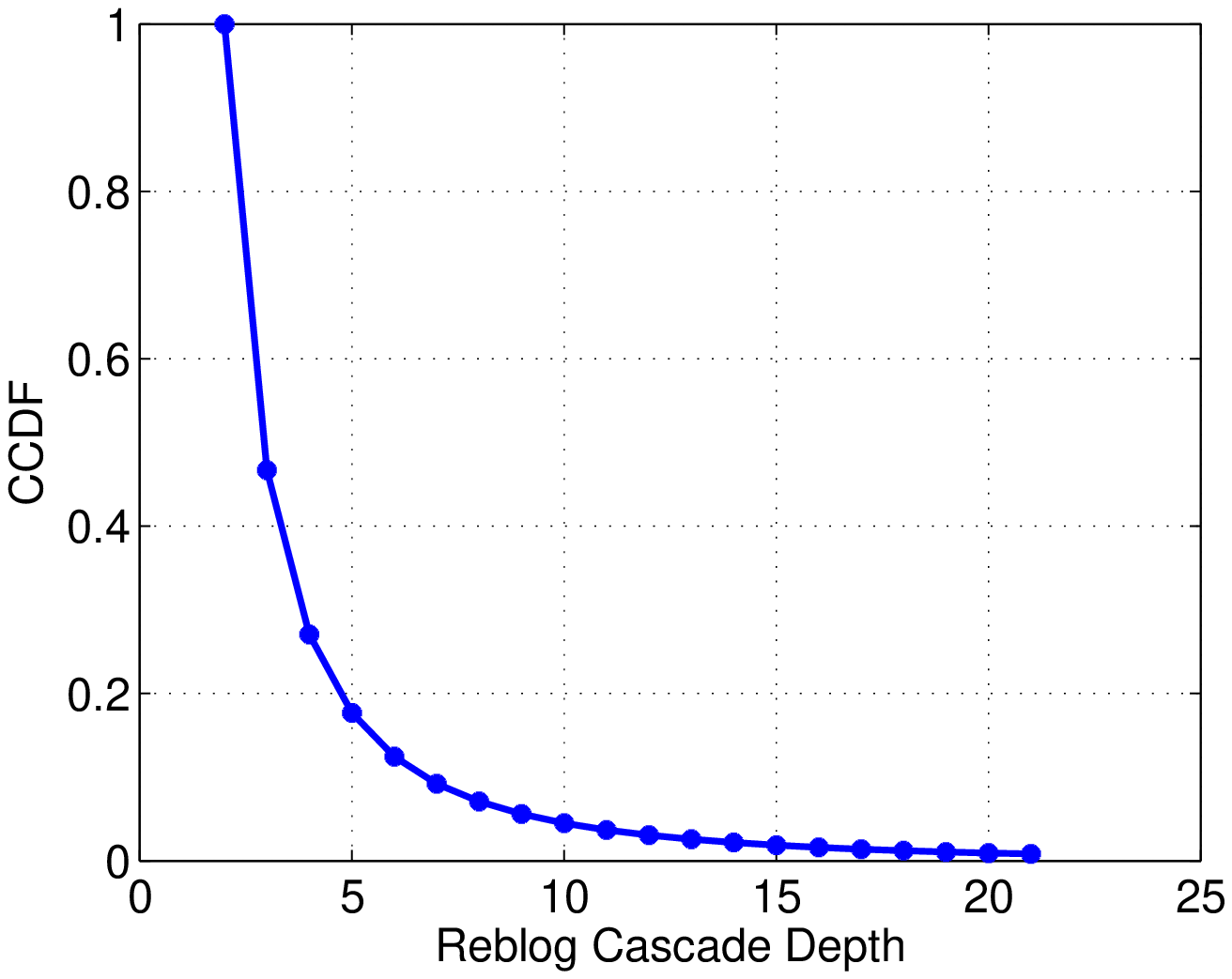,width=0.3\textwidth}
 \label{fig:tree-depth-ccdf-top20}}
\caption{Distribution of Reblog Cascade Depth}
  \label{fig:tree_depth}
\end{figure}

Figure~\ref{fig:tree_depth} plots the
distribution of number of hops: again, the reblog cascade depth distribution follows a power
law as well according to the PDF; when zooming into the CCDF,  we observe that only
$9.21\%$ of reblog cascades have depth larger than $6$. That is,
majority of cascades can reach just few hops, which is consistent with
the findings reported over Twitter~\cite{duncanWSDM2011}. Actually, $53.31\%$ of
cascades in Tumblr have depth $2$. Nevertheless, the maximum depth among all
cascades can reach 241 based on two week data.
This looks unlikely at first glimpse, considering any two users are
just few hops away.  Indeed, this is because users can add comment while reblogging, and
thus one user is likely to involve in one reblog cascade multiple
times. We notice that some Tumblr users adopt reblog as one way for
conversation or chat.

\begin{figure*}[t]
  \centering
  \includegraphics[width=\textwidth]{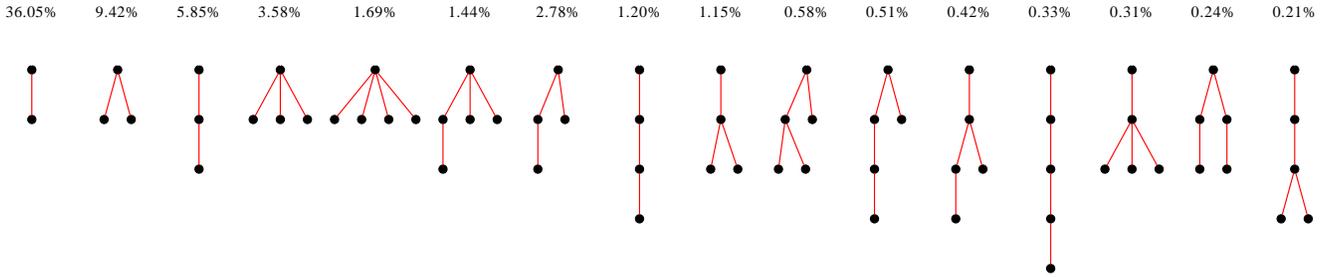}
  \caption{Cascade Structure Distribution up to  Size 5. The
    percentage at the top is the coverage of cascade structure.}
  \label{fig:structure}
\end{figure*}

{\bf Reblog Structure Distribution.}
Since most reblog cascades are few hops,  here we show the cascade
tree structure distribution up to size $5$ in
Figure~\ref{fig:structure}.  The structures are sorted based on their
coverage.  Apparently,  a substantial percentage of cascades ($36.05\%$) are of size $2$,  i.e.,
a post being reblogged  merely once.  Generally speaking, a reblog cascade of a flat
structure tends to have a higher probability than a reblog cascade of the same
size but with a deep structure. For instance, a reblog cascade of size 3 have
two variants, of which the flat one covers $9.42\%$ cascade while the
deep one drops to $5.85\%$.  The same patten applies to reblog cascades of size 4
and 5.   In other words, it is easier to spread a message
widely rather than deeply in general. This implies that it might be
acceptable to consider only the cascade effect under few hops and
focus those nodes with larger audience when one tries to maximize
influence or information propagation.

\begin{figure}
  \centering
  \includegraphics[width=0.3\textwidth]{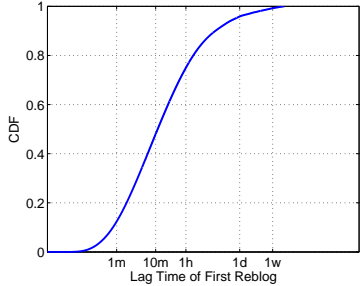}
  \caption{Distribution of Time Lag between a Blog and its first Reblog}
  \label{fig:time_lag}
\end{figure}

{\bf Temporal patten of reblog.} We have investigated the information
propagation spatially in terms of network topology, now we study how
fast for one blog to be reblogged?  Figure~\ref{fig:time_lag} displays
the distribution of time gap between a post and its first
reblog. There is a strong bias toward recency.
The larger the time gap since a blog is posted, the less
likely it would be reblogged.
$75.03\%$ of first reblog arrive within the first hour since a blog is
posted, and $95.84\%$ of first reblog appears within one day.
Comparatively,
It has been reported that ``half of retweeting occurs within an hour and $75\%$
under a day''~\cite{KwakWWW2010} on Twitter. In short, Tumblr reblog has a strong bias toward recency, and
information propagation on Tumblr is fast.


\section{Related Work}
\label{sec:relatedwork}

There are rich literatures on both existing and emerging online social
network services. Statistical patterns across different types of
social networks are reported, including traditional
blogosphere~\cite{ShiICWSM2007}, user-generated content platforms like
Flickr, Youtube and LiveJournal~\cite{Misl-etal07},
Twitter~\cite{Java2007,KwakWWW2010}, instant messenger
network~\cite{Lesk-Horv08}, Facebook~\cite{ugander2011anatomy}, and
Pinterest~\cite{Pinterest2013,ottoni2013ladies}.  Majority of them
observe shared patterns such as long tail distribution for user
degrees (power law or power law with exponential cut-off), small
($90\%$ quantile effective) diameter, positive degree association,
homophily effect in terms of user profiles (age or location), but not
with respect to gender.  Indeed, people are more likely to talk to the
opposite sex~\cite{Lesk-Horv08}. The recent study of
Pinterest observed that ladies tend to be more
active and engaged than men~\cite{ottoni2013ladies}, and women and men have different
interests~\cite{CSCW14Pinterest}.  We have compared Tumblr's patterns with
other social networks in Table~\ref{tab:network-stats} and
observed that most of those trend hold in Tumblr except for some
number difference.

%
Lampe \textit{et al.} ~\cite{facebook2007} did a set of survey
studies on Facebook users, and shown that people use Facebook to
maintain existing offline connections.  Java \textit{et al.}
~\cite{Java2007} presented one of the earliest research paper for
Twitter, and found that users leverage Twitter to talk their daily
activities and to seek or share information. In addition, Schwartz
~\cite{Pinterest2013} is one of the early studies on Pinterest, and
from a statistical point of view that female users repin more but
with fewer followers than male users. While Hochman and
Raz~\cite{Instagram2012} published an early paper using Instagram
data, and indicated differences in local color usage, cultural
production rate, for the analysis of location-based visual
information flows.

Existing studies on user influence are based on social networks or
content analysis. McGlohon \textit{et al.} ~\cite{jure2007ICWSM} found
topology features can help us distinguish blogs, the temporal activity
of blogs is very non-uniform and bursty, but it is
self-similar. Bakshy \textit{et al.} \cite{duncanWSDM2011}
investigated the attributes and relative influence based on Twitter
follower graph, and concluded that word-of-mouth diffusion can only be
harnessed reliably by targeting large numbers of potential
influencers, thereby capturing average effects. Hopcroft \textit{et
  al.} \cite{Hopcroft2011} studied the Twitter user influence based on
two-way reciprocal relationship prediction. Weng \textit{et al.}
\cite{Weng2010} extended PageRank algorithm to measure the influence
of Twitter users, and took both the topical similarity between users
and link structure into account. Kwak \textit{et al.}
\cite{KwakWWW2010} study the topological and geographical properties
on the entire Twittersphere and they observe some notable properties
of Twitter, such as a non-power-law follower distribution, a short
effective diameter, and low reciprocity, marking a deviation from
known characteristics of human social networks.

However, due to data access limitation, majority of the existing
scholar papers are based on either Twitter data or traditional
blogging data.  This work closes the gap by providing the first
overview of Tumblr so that others can leverage as a stepstone to
investigate more over this evolving social service or compare with
other related services.

\section{Conclusions and Future Work}

In this paper, we provide a statistical overview of Tumblr in terms of social network structure,
content generation and information propagation.
We show that Tumblr serves as a social
network, a blogosphere and social media simultaneously. It
provides high quality content with rich multimedia information,
which offers unique characteristics to attract youngsters.
Meanwhile, we also
summarize and offer  as rigorous comparison as possible with other social services
based on numbers reported in other papers.  Below we highlight some key
findings:
\begin{itemize}
\item With multimedia support in Tumblr, photos and text account for
  majority of blog posts, while audios and videos are still rare.
\item  Tumblr, though initially proposed for blogging,
yields a significantly different network structure from
traditional blogosphere. Tumblr's network is much denser and better connected. Close to $29.03\%$ of connections on Tumblr are
reciprocate, while blogosphere has only $3\%$.  The average distance
between two users in Tumblr is $4.7$, which is roughly half of that in
blogosphere.  The giant connected component covers $99.61\%$ of
nodes as compared to $75\%$ in blogosphere.
\item  Tumblr network is highly similar to
Twitter and Facebook, with power-law distribution for in-degree
distribution, non-power law out-degree distribution,  positive degree associativity for reciprocate
connections, small distance between connected nodes, and a dominant
giant connected component.
\item Without post length limitation,  Tumblr users tend to post
  longer. Approximately 1/4 of text posts have authentic contents beyond 140 bytes, implying
  a substantial portion of high quality blog posts for other tasks like topic
\item Those social celebrities tend to be more active. They post
  analysis and text mining.
  and reblog more frequently, serving as both content generators and
  information transmitters.   Moreover, frequent bloggers like to
  write short, while infrequent bloggers spend more effort in writing
  longer posts.
\item In terms of duration since registration, those long-term users and recently registered users post less
  frequently. Yet, long-term users reblog more.
\item  Majority of reblog cascades are
  tiny in terms of both size and depth, though extreme ones are
  not uncommon.   It is relatively easier to propagate a message wide
  but shallow rather than deep, suggesting the priority for influence
  maximization or information propagation.
\item Compared with Twitter, Tumblr is more vibrant and faster in
  terms of reblog and   interactions.  Tumblr reblog has a strong bias
  toward recency. Approximately $3/4$ of the first reblogs occur within the first hour
  and $95.84\%$ appear within one day.
\end{itemize}

This snapshot research is by no means to be complete. There are
several directions to extend this work.  First, some patterns
described here are correlations. They do not illustrate the underlying
mechanism.   It is imperative to differentiate correlation and
causality~\cite{Anag-etal08} so that we can better understand the user behavior.
Secondly, it is observed that
Tumblr is very popular
among young users, as half of Tumblr's visitor base being under 25
years old. Why is it so?
We need to combine
content analysis, social network analysis, together with user profiles
to figure out. In addition, since more than $70\%$ of Tumblr posts
are images, it is necessary to go beyond photo captions,
and analyze image content together with other meta information.

\bibliographystyle{aaai} \bibliography{reference,tumblr}

\end{document}